# Breaking the Trade-off: Bulk 2D Ising Superconductivity with High $T_c$ and Giant Interlayer Spacing via a Unique Chain Intercalation in $(BaS)_{1/3}TaS_2$

Ziyi Zhu,[1,†] Leiming Chen,[1,*] Xiangqi Liu,[2,†] Haonan Wang,[3,†] Chen Xu,[2,†] Ze Yan,[2] Zhengyang Li,[1] Wei Xia,[2,4] Jiawei Luo,[2,4] Na Yu,[2,5] Xia Wang,[2,5] Ke Qu,[3] Zhenzhong Yang[3,*] Yanfeng Guo[2,4,*]

[1]*Henan Key Laboratory of Aeronautical Materials and Application Technology, Zhengzhou University of Aeronautics, Zhengzhou, Henan 450046, China*

[2]*State Key Laboratory of Quantum Functional Materials, School of Physical Science and Technology, ShanghaiTech University, Shanghai 201210, China*

[3]*Key Laboratory of Polar Materials and Devices (MOE), Department of Electronics, School of Information and Electronic Engineering, East China Normal University, Shanghai 200241, China.*

[4]*ShanghaiTech Laboratory for Topological Physics, ShanghaiTech University, Shanghai 201210, China*

[5]*Analytical Instrumentation Center, School of Physical Science and Technology, ShanghaiTech University, Shanghai 201210, China*

[†]These authors contributed equally to this work.

*Corresponding authors: lmchen@zua.edu.cn; zzyang@phy.ecnu.edu.cn;
guoyf@shanghaitech.edu.cn.

**ABSTRACT**

**Two-dimensional (2D) transition-metal dichalcogenides (TMDs) are promising platforms for low-dimensional superconductivity. However, in conventional intercalated systems, achieving a high superconducting transition temperature ($T_c$) often comes at the expense of reduced interlayer spacing and weakened 2D character. Here, we overcome this long-standing compromise through a unique chain-like intercalation strategy. We report the synthesis and properties of a new polymorph, $(BaS)_{1/3}TaS_2$, in which a distinctive Ba-S-S-Ba chain structure is inserted between $TaS_2$ bilayers. This unique configuration breaks the bulk *c*-axis mirror symmetry while achieving exceptional interlayer decoupling, with an inter-bilayer spacing of 12.75 Å—more than three times that of pristine 2H-$TaS_2$. By suppressing interlayer electronic coupling, this structural evolution allows local inversion symmetry breaking within individual $TaS_2$ layers to dominate. This prevents compensation of the Ising spin–orbit fields typical of centrosymmetric bulk phases, enabling robust 2D Ising superconductivity. Remarkably, the compound exhibits an enhanced $T_c$ without sacrificing its large interlayer spacing, thereby breaking the conventional trade-off between "large spacing/high anisotropy" and "high $T_c$". Comprehensive transport, magnetic, and thermodynamic measurements confirm its robust superconducting state. Our work establishes a versatile intercalation framework for designing bulk-like 2D Ising superconductors, providing a new route to reconcile competing material demands and expanding the scope of Ising superconductivity research.**

## INTRODUCTION

Transition-metal dichalcogenides (TMDs), with their archetypal $MX_2$ (M represents a transition metal and X denotes a chalcogen element) layered structure (Figure 1(a)), have established themselves as a paradigmatic platform for exploring low-dimensional superconductivity. In these systems, the intricate interplay among dimensionality, spin-orbit coupling (SOC), and symmetry breaking gives rise to exotic quantum states, notably Ising superconductivity.[1-12] Monolayer TMDs inherently lack inversion symmetry, which induces a robust out-of-plane spin texture—Ising SOC that protects Cooper pairs from in-plane magnetic fields and stabilizes two-dimensional (2D) superconducting order.[4-6] Further symmetry engineering, such as through a non-periodic chemical potential or tailored superlattices (Figure 1(b)), can break the *c*-axis mirror symmetry by inducing a local out-of-plane potential gradient, thereby introducing complementary Rashba-type SOC components.[13] These symmetry-controlled electronic structures are vividly illustrated in the layer-dependent Fermi surface evolution of $TaS_2$ (Figure 1(c-d)): in the monolayer limit, spins at the K and K' valleys are locked perpendicular to the *ab*-plane, stabilized by a strong Ising field $B_{so}$, which is illustrated by the dashed line along K-Γ-K' of the Brillouin zone in the electronic band structure. In contrast, in bulk or multilayer 2H-phase TMDs, finite interlayer Josephson coupling which is characterized by a non-zero interlayer transfer integral $t^{\perp}$ leads to compensation of Ising fields between adjacent layers, significantly weakening both 2D confinement and Ising protection—thereby obstructing the realization of robust, clean 2D superconductivity in bulk forms.[4,6]

To tailor these superconducting properties, intercalation has emerged as a widely used strategy that exploits the weak van der Waals (vdW) gaps between TMD layers.[14-20] However, conventional intercalation methods face a persistent trade-off: small ionic intercalants (e.g., $Na^+$,

K$^+$) effectively raise the superconducting transition temperature ($T_c$) by enhancing the density of states (DOS) at the Fermi level, but simultaneously strengthen interlayer coupling, driving a dimensional crossover to 3D behavior and undermining Ising SOC, seen in Figure 1(e).[21-39] On the other hand, bulky organic or inorganic intercalants can expand the interlayer spacing and recover 2D character, yet their electronic inertness typically suppresses $T_c$ to impractically low values, seen in Figure 1(f), limiting experimental studies of underlying pairing mechanisms.[18-21] While tailored organic intercalation has successfully maintained high $T_c$ in systems such as molecule-intercalated $NbSe_2$ and $TaS_2$, these methods often based on surface-limited electrochemical or liquid-phase ion exchange, which can suffer from non-uniform intercalation and lattice distortions.[40-42] Achieving both high $T_c$ and robust 2D Ising superconductivity in a homogenous bulk system thus remains a central challenge—one whose resolution would open the door to investigating delicate quantum phenomena in macroscopic, readily accessible samples.[26, 27]

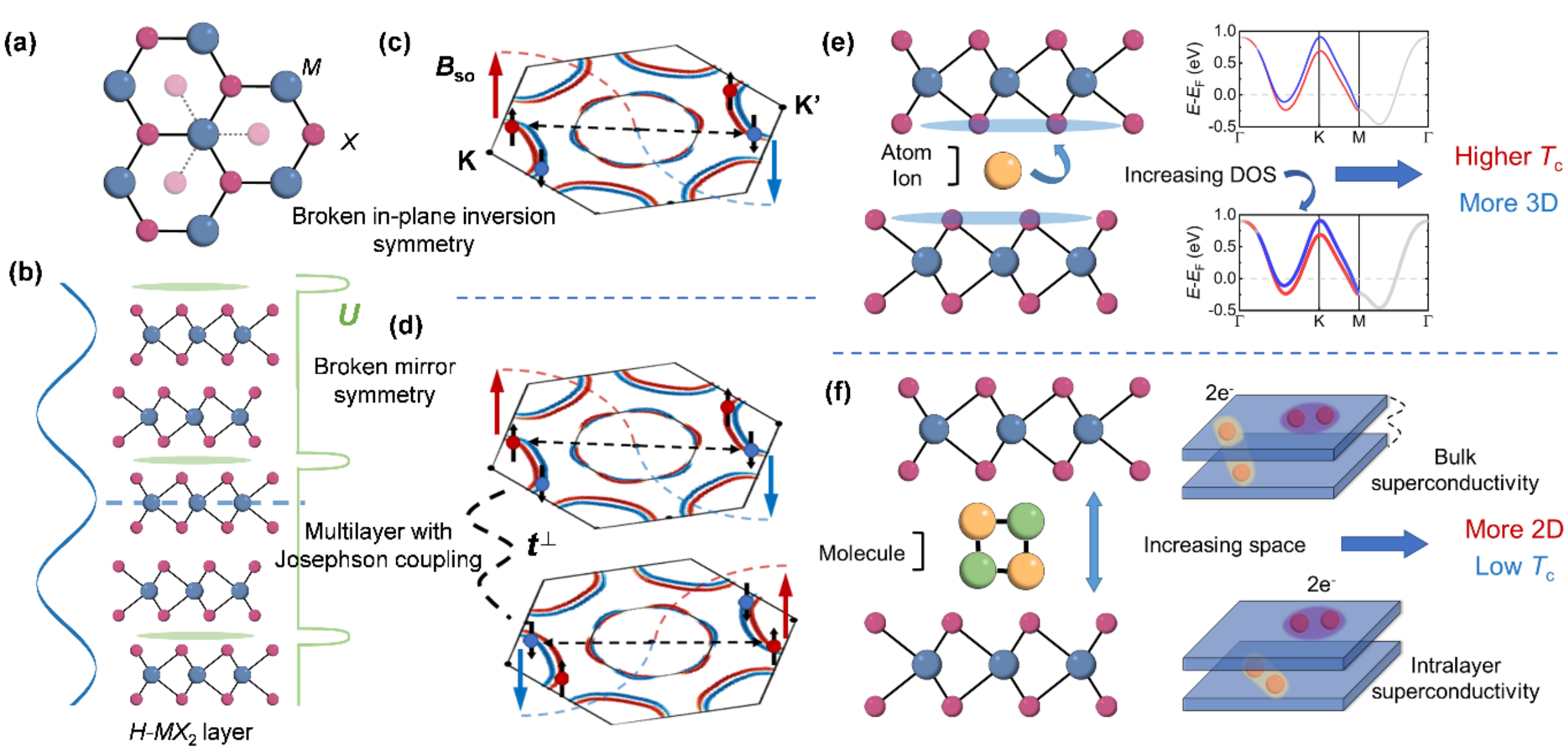

**Figure 1.** Symmetry breaking, SOC, and intercalation in TMDs. (a) Top view of the H-$MX_2$ structure (M: blue; X: red). Inversion symmetry is broken within each layer due to missing inversion-related X-atom sites. (b) The broken *c*-axis mirror symmetry induces an out-of-plane

potential gradient, adding a Rashba-type SOC component. (c) Schematic of monolayer H-$TaS_2$: spins in the K and K' valleys are pinned out-of-plane by the Ising SOC field $B_{so}$. Color-coded arrows depict spin projections; the dashed envelope along K-Γ-K' indicates the $B_{so}$ distribution. (d) Schematic Fermi surface of multilayer $TaS_2$ with finite interlayer coupling $t^{\perp}$. Opposite $B_{so}$ orientations in adjacent layers lead to partial cancellation of Ising SOC. (e) Small-radius intercalants increase the DOS at the Fermi level, raising $T_c$, but also enhance interlayer Josephson coupling. This drives a dimensional crossover from 2D to 3D and weakens Ising protection. (f) Bulky intercalants expand the interlayer spacing, restoring monolayer-like 2D superconductivity, but typically suppress $T_c$ to very low values.

Recent advances in structural engineering via intercalation have shown promise in tuning the dimensionality of TMD-based superconductors. Examples include bilayer Sn-intercalated $TaSe_2$ and block layer-intercalated $NbS_2$ superlattices, which exhibit modified interlayer coupling.[19-22] However, these systems still face inherent limitations: the former retains strong interlayer coupling, leading to 3D superconductivity, while the latter suffers from a significantly suppressed $T_c$. Although Ising superconductivity has been realized in some bulk TMDs, existing systems generally fail to achieve a synergistic optimization of high $T_c$ and robust 2D anisotropy. This underscores a critical need for a novel intercalation strategy that can simultaneously decouple adjacent TMD layers and provide effective charge doping—thereby unifying high-Tc superconductivity with strong 2D Ising behavior in a bulk material.

Here, we address this challenge by introducing a novel chain-like intercalation architecture in $(BaS)_{1/3}TaS_2$, a newly synthesized polymorph that hosts bulk 2D Ising superconductivity. Unlike conventional single-layer intercalation motifs (e.g., PbS-intercalated single-layer $TaS_2$),[35] the unique Ba-S-S-Ba chain structure inserted between bilayer $TaS_2$ not only breaks the *c*-axis mirror symmetry of the bulk crystal but also substantially expands the interlayer spacing—effectively decoupling neighboring $TaS_2$ layers while maintaining electronic activity. Notably, in contrast to previously reported bilayer-intercalated systems where significant interlayer coupling persists (e.g., 2Sn-2$TaSe_2$, where bilayer Sn intercalation retains significant Josephson coupling and drives 3D superconductivity),[22] $(BaS)_{1/3}TaS_2$ achieves an optimal synergy between doping-induced $T_c$ enhancement and interlayer-decoupling-enhanced 2D confinement.

Through comprehensive transport, magnetic, and thermodynamic measurements, we demonstrate that $(BaS)_{1/3}TaS_2$ exhibits definitive signatures of Ising superconductivity, including an elevated $T_c$, an in-plane upper critical field $B_{c2}^{//}$ surpassing the Pauli limit $B_P$, and pronounced superconducting anisotropy. Systematic comparison with other superconducting TMDs confirms that our chain-like intercalation strategy successfully overcomes the conventional trade-off between $T_c$ and dimensionality. This work establishes a general materials design pathway toward bulk 2D Ising superconductors with optimized properties, providing a versatile platform for exploring topologically non-trivial superconducting states and enabling future device applications.

## EXPERIMENTAL SECTION

**Synthesis of $(BaS)_{1/3}TaS_2$ single crystals.** High-quality $(BaS)_{1/3}TaS_2$ single crystals were synthesized using a flux method. High-purity BaS (99.5%), Ta powder (99.9%), S powder (99.999%), and anhydrous $BaCl_2$ powder (99.9%) were mixed with the molar ratio of 0.5:1:2:1 in a glove box and placed into an evacuated quartz tube, which was then sealed. The mixture was heated to 1373 K over a period of 25 hours, the furnace was slowly cooled down to 800°C in 7 days, followed by a power-off cooling. Single crystals with typical dimensions of 0.5 × 0.5 × 0.05 $mm^3$ were obtained. These crystals were found to be stable, showing no degradation or structural phase transitions when exposed in air.

**Crystal characterizations.** The composition of the $(BaS)_{1/3}TaS_2$ single crystals was analyzed using a Phenom Pro scanning electron microscope (SEM) equipped with energy-dispersive X-ray spectroscopy (EDS). Room-temperature powder X-ray diffraction (PXRD) patterns were collected from crushed single-crystal samples using a Bruker D8 Venture diffractometer with Cu $K_\alpha$ radiation ($\lambda = 1.5418$ Å). Single-crystal X-ray diffraction (SXRD) measurements were performed at room temperature on a Bruker D8 diffractometer with Mo $K_\alpha$ radiation ($\lambda = 0.71073$ Å).

**Magnetization and electrical transport measurements.** Magnetic characterization was carried out using a Quantum Design MPMS-3 magnetometer. Temperature-dependent magnetization $M(T)$ was measured under a small field (5-20 Oe) applied perpendicular to the *ab*-plane in both zero-field-cooling (ZFC) and field-cooling (FC) modes. Isothermal magnetization curves were acquired over a field range of ±600 Oe at various temperatures. Electrical transport measurements were performed using a standard four-wire method in a Quantum Design DynaCool Physical Property Measurement System (PPMS-14T) equipped with a dilution refrigerator. The *c*-axis resistance was measured by the four-electrode method with the Corbino-shape-like configuration.[13] The specific heat was measured using the standard thermal-relaxation technique in the PPMS system, with careful background subtraction over the temperature range from 300 K down to 1.8 K.

**Cs-corrected scanning transmission electron microscopy measurements.** Cross-sectional transmission electron microscopy (TEM) samples were prepared with a Thermo Fisher Helios G4 UX focused-ion-beam scanning electron microscope (FIB-SEM). Atomic-resolution high-angle annular dark-field (HAADF) scanning transmission electron microscopy (STEM) images were acquired using a JEOL JEM-ARM300F Cs-corrected STEM operated at 300 kV.

## RESULTS AND DISCUSSION

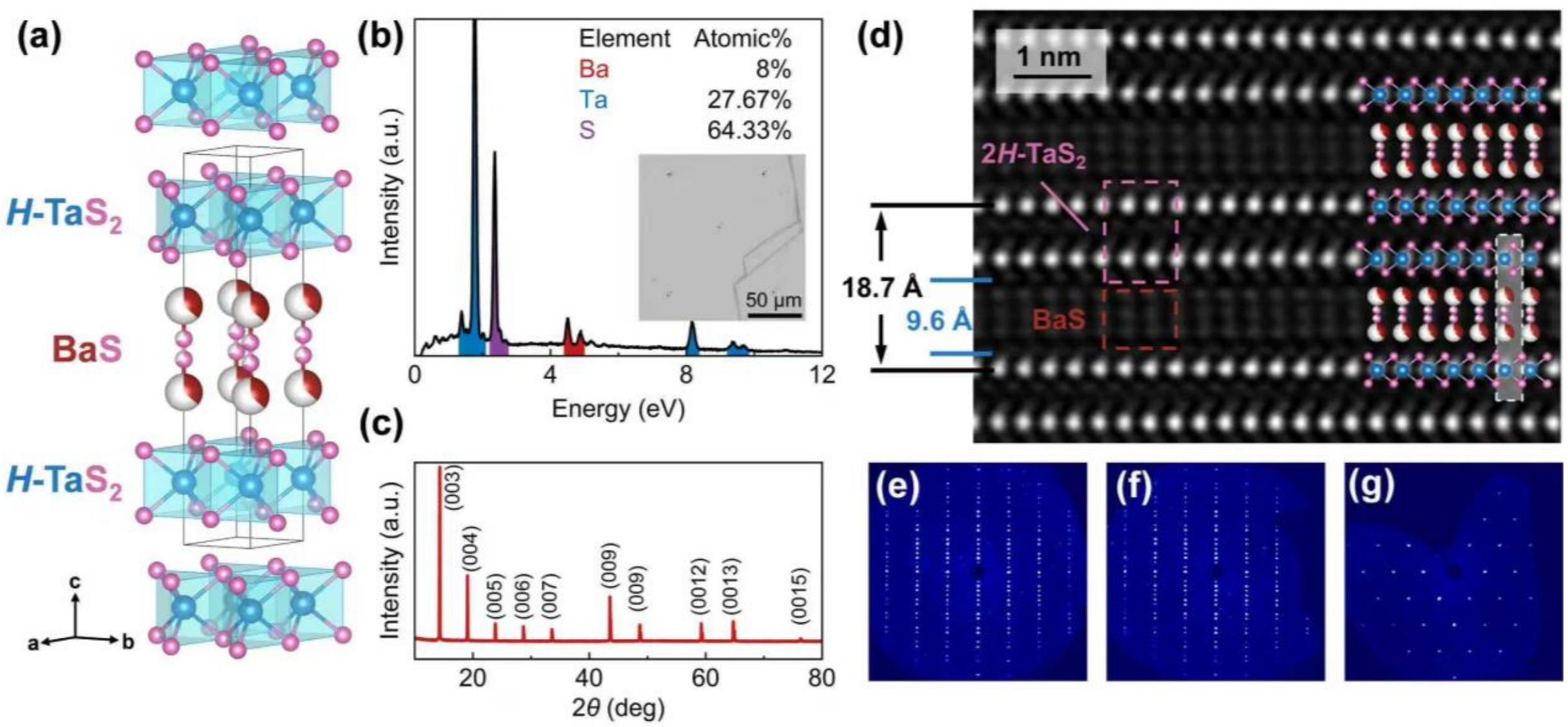

**Figure 2.** Crystal structure and structural characterization of $(BaS)_{1/3}TaS_2$. (a) The crystal structure of $(BaS)_{1/3}TaS_2$, distinct from pristine 2H-$TaS_2$. Ba-S-S-Ba chains are intercalated with a two-layer periodicity, forming locally decoupled $TaS_2$ bilayers. (b) EDS spectrum confirming the stoichiometric molar ratio Ba:Ta:S ~ 0.33:1:2.33. (c) XRD pattern exhibiting sharp (00*l*) reflections, consistent with high crystallinity. (d) HAADF-STEM image with the corresponding structural model overlaid. The unit cell ($a$ = $b$ = 3.32 Å, $c$ = 18.74 Å) contains inversion-related H-$TaS_2$ bilayers, whose mirror symmetry is broken by the adjacent block layers. The inter-bilayer spacing ($d$ = 9.6 Å) exceeds three times that of 2H-$TaS_2$, strongly suppressing the interlayer transfer integral $t^{\perp}$ and enhancing the 2D character of the electronic structure. (e-g) SXRD patterns in reciprocal space, viewed along the (*0kl*), (*h0l*), and (*hk0*) planes, respectively.

Figure 2(a) illustrates the unique crystal structure of $(BaS)_{1/3}TaS_2$, which crystallizes in the space group $P$-3$m$1—a notable departure from the $P6_3/mmc$ symmetry of pristine 2H-$TaS_2$. The structure features a two-layer periodic insertion of Ba-S-S-Ba chains, forming a locally decoupled bilayer architecture. EDS confirms the stoichiometry, with a Ba:Ta:S molar ratio of approximately 0.33:1:2.33 (Figure 2(b)). The high crystallinity of as-grown $(BaS)_{1/3}TaS_2$ is evidenced by sharp (00*l*) Bragg peaks in PXRD (Figure 2(c)) and well-defined diffraction spots from SXRD reconstructions (Figure 2(e-g)). A cross-sectional HAADF-STEM image with an overlaid structural model further visualizes the layered arrangement (Figure 2(d)).

Refined both PXRD and SXRD analyses yield a unit cell (space group $P$-3$m$1, $a$ = 3.32 Å, $c$ = 18.74 Å) comprising an inversion-related H-$TaS_2$ bilayer (see details in Supporting Information). While the overall unit cell retains inversion symmetry, the adjacent block layers break mirror symmetry across the bilayer. The inter-bilayer spacing ($d \approx 9.6$ Å) is more than three times larger than that of pristine 2H-$TaS_2$,[36] leading to a drastic suppression of the interlayer transfer integral $t^{\perp}$. This pronounced expansion of the vdW gap strongly enhances the 2D character of the electronic structure. By effectively suppressing the interlayer hopping, the system approaches a quasi-monolayer limit, which provides a structural basis for preserving the strong spin-orbit coupling inherent to the individual non-centrosymmetric layers.

Figure 3(a) shows the resistivity-temperature ($R$-$T$) profile of a $(BaS)_{1/3}TaS_2$ single crystal. The absence of the characteristic charge-density-wave (CDW) transition of 2H-$TaS_2$ near 75 K indicates that intercalation suppresses the CDW state. In the high-temperature range (300–50 K), the normal-state resistivity follows a nearly linear power-law behavior $R(T) = cT^{\alpha} + R_0$, with $\alpha \approx 0.93$—close to unity and indicative of strange-metal character. Between 10 and 50 K, $\alpha$ systematically increases toward a Fermi-liquid-like regime ($\alpha \approx 1.55$), approaching the ideal Fermi liquid exponent ($\alpha = 2$). This evolution suggests a link between enhanced superconductivity and non-Fermi-liquid (strange-metal) behavior in intercalated TMDs.[37] Below ~ 3 K, the sample enters into the superconducting state, achieving zero resistance at $T_{c0} = 3.1$ K—significantly higher than that of pristine 2H-$TaS_2$ ($T_c \sim 1.0$ K). The marked enhancement of $T_c$ in $(BaS)_{1/3}TaS_2$ arises from the synergistic action of CDW suppression and charge transfer from the intercalated Ba-S-S-Ba chains to the $TaS_2$ host lattice.[6,19] The elimination of the competing CDW order relieves the occupation competition for electronic states at the Fermi level, fundamentally removing a key bottleneck for Cooper pair formation, while the electron doping induced by interlayer charge transfer elevates the carrier density and density of states at the Fermi level, which further reinforces superconducting pairing interactions. The expanded interlayer spacing inherently strengthens the 2D character of the electronic system, leading to pronounced electrical anisotropy. Figure 3(b) displays the resistivity anisotropy $\rho_{zz}/\rho_{xx}$ measured between 300 and 5 K, with $\rho_{zz}$ obtained using a Corbino geometry.[13] The ratio rises sharply upon cooling, exceeding $4 \times 10^3$ at low temperatures. The non-metallic interlayer transport indicates a strongly 2D Fermi surface; beyond conventional Boltzmann relaxation, interlayer hopping via resonant tunneling through impurity levels (e.g., Ba/S vacancies in spacer layers) must also be considered.

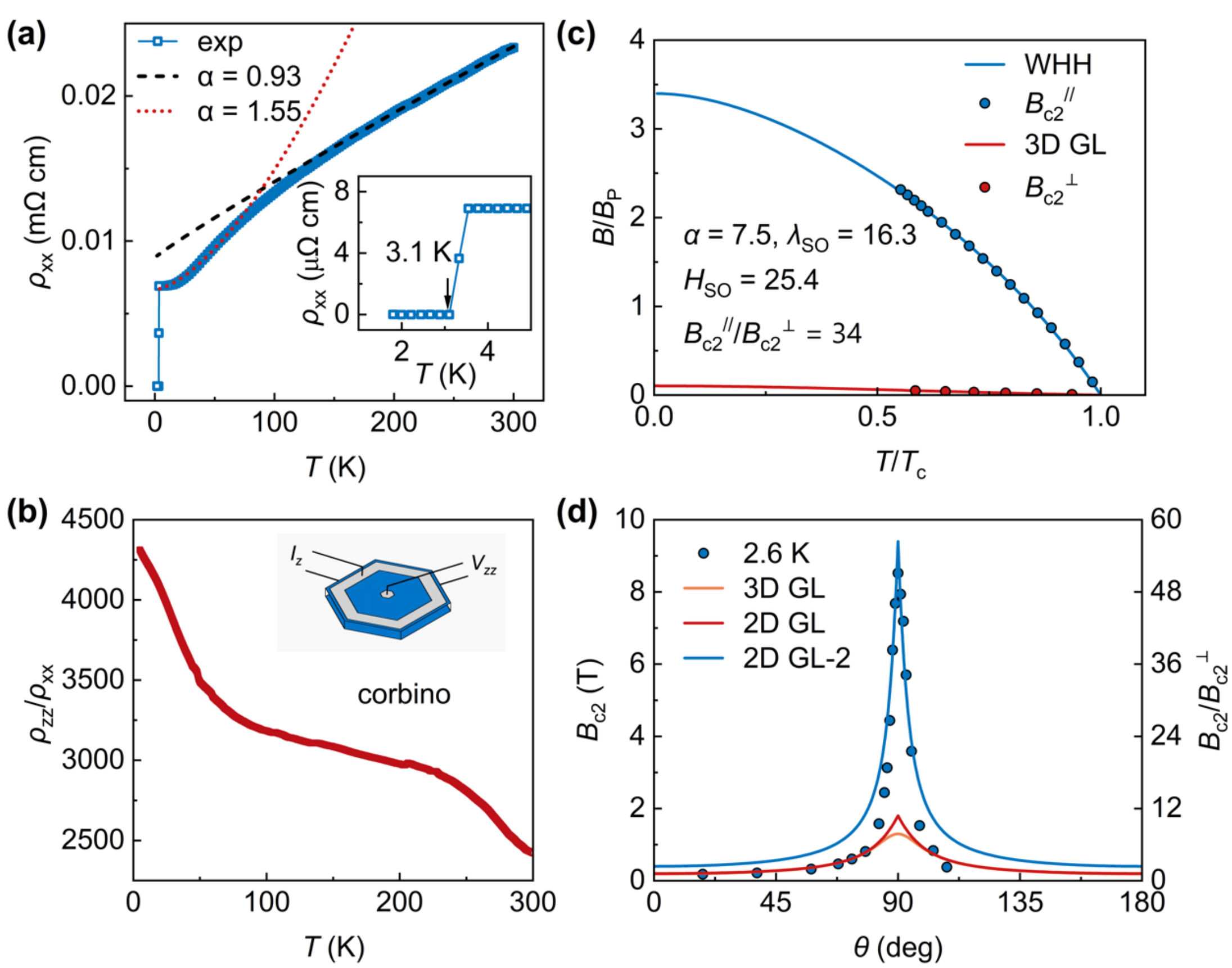

**Figure 3.** Transport properties and upper critical field of $(BaS)_{1/3}TaS_2$ single crystal. (a) Temperature dependence of resistivity of $(BaS)_{1/3}TaS_2$. High-temperature (300-50 K) behavior follows $R(T) = cT^{\alpha} + R_0$ with $\alpha = 0.93$ (strange-metal-like), transitioning to Fermi-liquid~~-like~~ behavior ($\alpha = 1.55$) at 10-50 K, and a superconducting transition at $T_{c0} = 3.1$ K. (b) Resistivity anisotropy $\rho_{zz}/\rho_{xx}$ increasing to $4\times10^3$ at low temperatures, signaling strong 2D character. (c) Out-of-plane upper critical field $B_{c2}{}^{\perp}$-$T$ fits the 3D GL relation, while $B_{c2}{}^{/\!/}$-T is well described by the WHH model, yielding $B_{c2}{}^{/\!/}(0) = 20.4$ T (surpasses the weak-coupling $B_P$ by nearly a factor of four). (d) Angular dependence of $B_{c2}(\theta)$ at 2.6 K: two Tinkham fits are required ($|\theta - 90°| > 10°$ and $< 10°$), showing wide-range enhancement of in-plane $B_{c2}$ associated with Josephson vortex dynamics.

To further characterize the superconducting state, we measured the upper critical field parallel ($B_{c2}{}^{/\!/}$) and perpendicular ($B_{c2}{}^{\perp}$) to the layers. As shown in Figure 3(c), $B_{c2}{}^{\perp}(T)$ follows a linear dependence well described by the 3D Ginzburg-Landau (GL) relation (see Supporting Information), giving a zero-temperature GL coherence length $\xi^{\perp} = 29$ nm. In contrast, $B_{c2}{}^{/\!/}(T)$ exhibits a nonlinear trend incompatible with GL theory; instead, the data are accurately modeled by the Werthamer-Helfand-Hohenberg (WHH) formalism—a microscopic framework appropriate for dirty superconductors dominated by spin-orbit scattering (see Supporting Information). The WHH fit yields a zero-temperature in-plane upper critical field $B_{c2}{}^{/\!/}(0) = 20.4$ T, which exceeds the weak-coupling Pauli limit $B_P$ by nearly a factor of four.

Figure 3(d) presents the angular dependence of $B_{c2}(\theta)$ measured at 2.6 K (~ 0.8 $T_{c0}$). Intriguingly, for near-in-plane orientations ($|\theta - 90°| < 10°$), $B_{c2}(\theta)$ shows an extra enhancement beyond the prediction of standard 2D GL theory. The angular profile is described by two distinct Tinkham fits: for $|\theta - 90°| > 10°$, $B_{c2}{}^{/\!/} = 1.8$ T and $B_{c2}{}^{\perp} = 0.2$ T; for $|\theta - 90°| < 10°$, the values become $B_{c2}{}^{/\!/} = 9.4$ T and $B_{c2}{}^{\perp} = 0.4$ T. This broad-angle enhancement contrasts sharply with the narrow angular range ($|\theta - 90°| < 2°$) reported for thick 2H-$NbSe_2$ or other intercalated TMD superconductors.[14-19] The extended angular robustness of the enhanced $B_{c2}^{//}$ suggests that the 2D superconductivity arises not only from interlayer decoupling but also from Josephson vortex dynamics.[19]

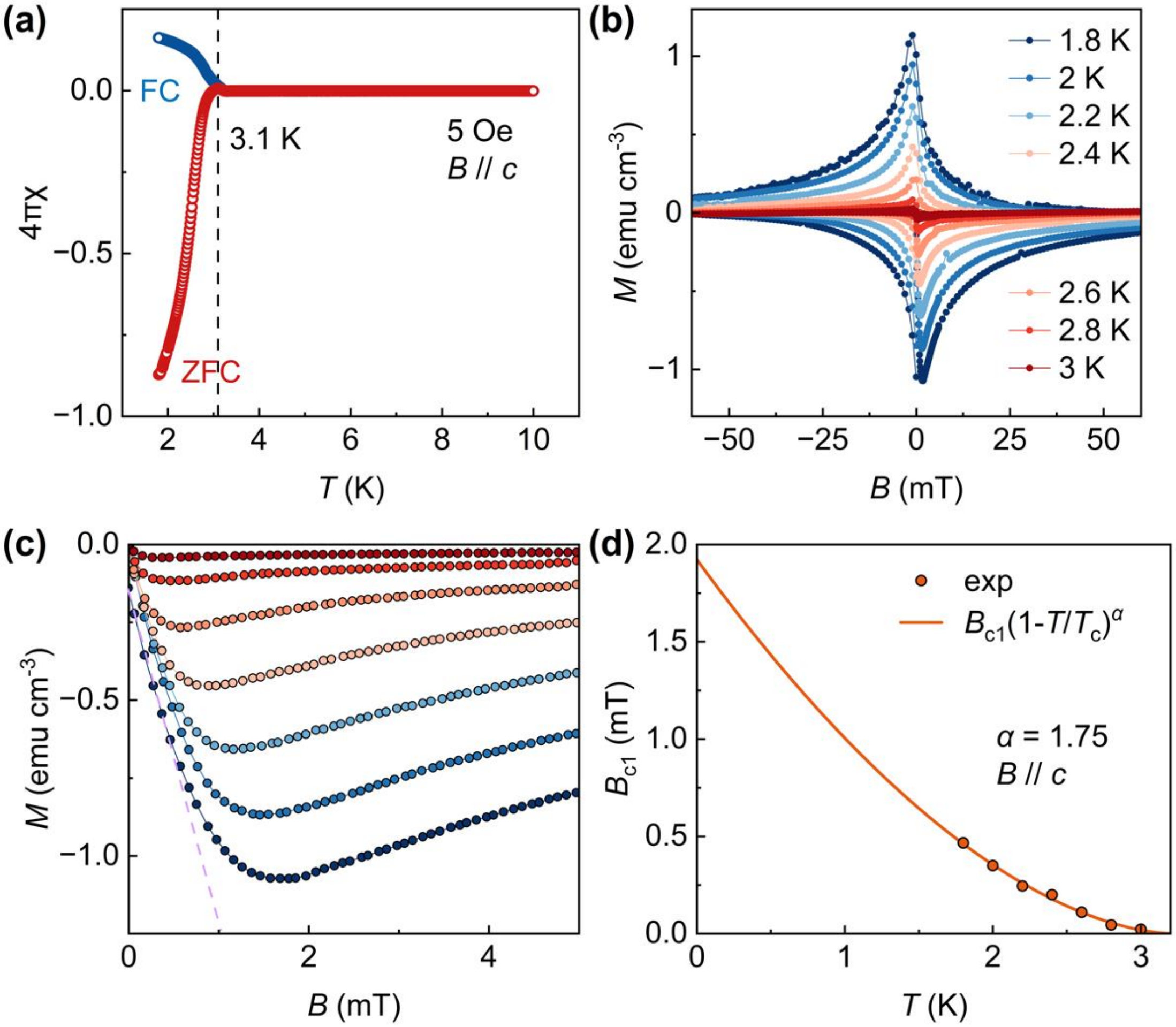

**Figure 4.** Magnetic properties of $(BaS)_{1/3}TaS_2$ single crystal. (a) ZFC and FC magnetic susceptibility curves, with separation arising from vortex pinning. (b) Superconducting magnetization hysteresis loops at different temperatures; critical current density $J_c$ (calculated via Bean model) reaches $1.2 \times 10^4$ A/cm$^2$ at 1.8 K, comparable to pristine TMD superconductors. (c) Field dependence of magnetization with field perpendicular to the crystal plane. The lower critical field $B_{c1}^{\perp} < 20$ Oe (estimated via $B_{c1}(T) = B_{c1}(0)(1-T/T_c)^{\alpha}$.

Figure 4(a) displays magnetic susceptibility measurements under ZFC and FC conditions; the separation between the curves reflects vortex pinning. The superconducting volume fraction, determined from the low-field linear slope after careful correction for the demagnetization factor (Supporting Information), reaches 87.1% at 1.8 K, confirming bulk superconductivity. Figure 4(b) shows magnetization hysteresis loops at selected temperatures. Using the Bean critical-state model, $J_c = 20\times\Delta M / [a\ (1 - a/3b)]$, where $\Delta M$ is the hysteresis width and a, b are sample dimensions, we obtain a critical current density $J_c \approx 1.2 \times 10^4$ A/cm$^2$ at 1.8 K—comparable to values reported for pristine TMD superconductors.[38, 39] Figure 4(c) plots magnetization versus field for $B \parallel c$. In the Meissner-shielding regime, the diamagnetic response varies linearly with field. Fitting the temperature dependence of the lower critical field with the phenomenological form $B_{c1}(T) = B_{c1}(0)(1-T/T_c)^{\alpha}$ (Figure 4(d)) yields an out-of-plane lower critical field $B_{c1}{}^{\perp} < 20$ Oe—similar to the behavior observed in Ising superconductor $4H_a$-$NbSe_2$.[39]

Specific heat measurements (Figure 5(a)) provide complementary thermodynamic evidence for bulk superconductivity. As shown in Figure 5(b), the zero-field data plotted as $C/T$ versus $T$ exhibit a clear anomaly near $T_c \approx 3.1$ K, consistent with transport and magnetic results. Fitting the normal-state data using the extended Debye model $C/T = \gamma_n + \beta T^2 + \eta T^4$ yields the electronic specific heat coefficient $\gamma_n$ (Figure 5(c)). The normalized specific heat jump $\Delta C/\gamma_n T_c \approx 1.42$ agrees well with the BCS value of 1.43. Moreover, the electronic specific heat is well described by a single *s*-wave gap model (Figure 5(d)). Together, these comprehensive transport, magnetic, and thermodynamic measurements firmly establish the realization of bulk two-dimensional superconductivity in $(BaS)_{1/3}TaS_2$ single crystals.

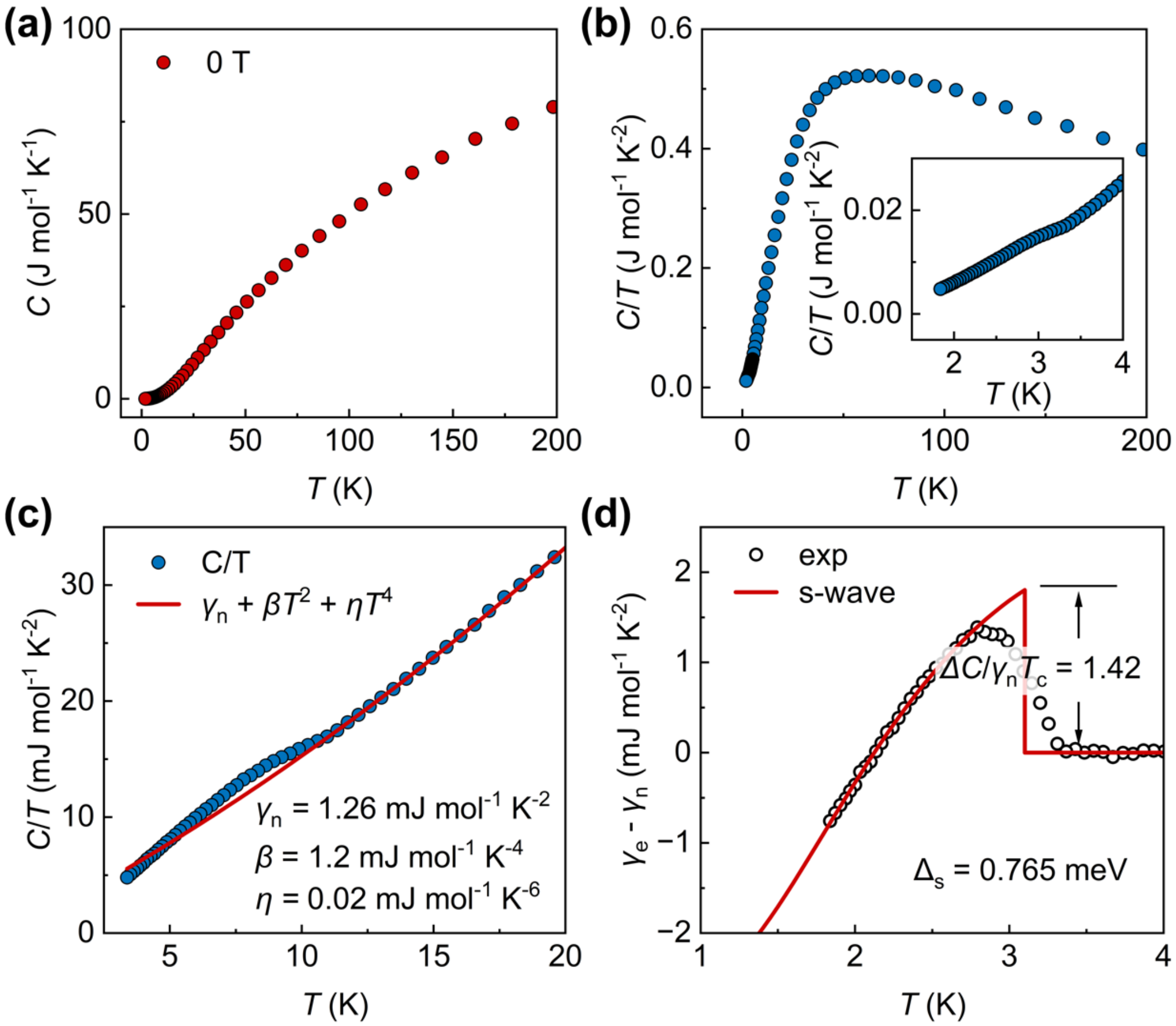

**Figure 5.** Specific heat of $(BaS)_{1/3}TaS_2$. (a) Raw data of $C(T)$ for $(BaS)_{1/3}TaS_2$ single crystals at 0 T. (b) Temperature-dependent heat capacity data. The superconducting transition is demonstrated by the inset of low-temperature data. (c) The normal state fitting curve by using the formula $C/T = \gamma_n + \beta T^2 + \eta T^4$. The obtained parameters are $\gamma_n$ = 1.26 mJ mol$^{-1}$ K$^{-2}$, $\beta$ = 1.2 mJ mol$^{-1}$K$^{-4}$ and $\eta$ = 0.02 mJ mol$^{-1}$K$^{-6}$. (d) The electronic specific heat data (circles) with theoretical fits to the BCS model by using the conventional *s*-wave. The $\Delta C/\gamma_n T_c \sim 1.42$ is close to the weak coupling BCS value of 1.43.

Figure 6 systematically compares key physical parameters across various TMD-based superconductors, visually highlighting the distinctive advantage of our work in balancing high $T_c$ and robust 2D superconductivity. Figure 6(a) presents the relationship between $T_c$ and $B_{c2}^{//}$. While $B_{c2}^{//}$ generally scales with $T_c$ across most TMD systems, simultaneous optimization of both parameters remains rare. The parent compound 2H-$TaS_2$ shows modest performance ($T_c \approx 0.9$ K, $B_{c2}^{//} \approx 0.54$ T). Alkali-metal-intercalated variants such as $Na_{0.1}TaS_2$ ($T_c = 4.3$ K, $B_{c2}^{//} = 16$ T)[28] and $Ni_{0.04}TaS_2$ ($T_c = 4.15$ K, $B_{c2}^{//} = 17.3$ T)[23] achieve higher $T_c$ through charge doping, yet their upper critical fields remain limited. Even high-$T_c$ systems like 2H-$NbSe_2$ ($T_c = 7.2$ K) only reach $B_{c2}^{//} \approx$ 15 T, falling short of the ideal range for strong Ising superconductivity. In striking contrast,

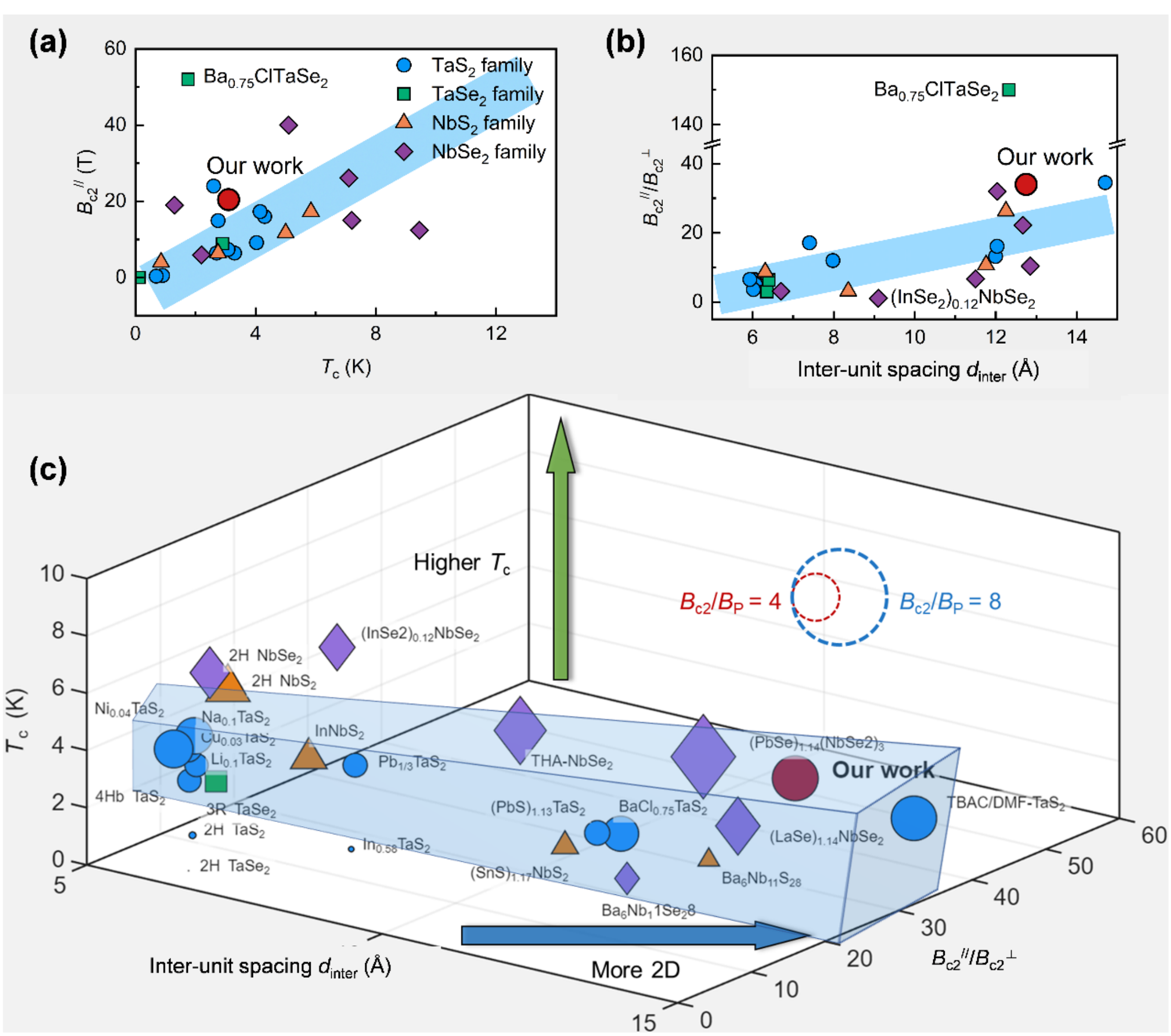

**Figure 6.** Correlations between key parameters of TMD-based superconducting systems. (a) $T_c$ versus $B_{c2}^{//}$. Most TMD systems show a positive correlation between these two quantities. (b) Inter-

$(MX_2)_n$-unit spacing $d_{inter}$ plotted against superconducting anisotropy $B_{c2}^{/\!/}/B_{c2}^{\perp}$. A clear positive trend is observed: expanded interlayer spacing suppresses the interlayer Josephson coupling $t^{\perp}$, a key structural factor promoting 2D superconducting character. (c) 3D scatter plot displaying anisotropy (*x*-axis), inter-unit spacing (*y*-axis), and $T_c$ (*z*-axis). Symbol size scales with the Ising SOC strength $B_{c2}^{/\!/}/B_P$. Most systems cluster either in the "low anisotropy–small spacing–low $T_c$" region (parent TMDs) or the "high anisotropy–large spacing–low $T_c$" region (macromolecule-intercalated compounds). In contrast, $(BaS)_{1/3}TaS_2$ occupies a distinct region characterized by high anisotropy (34), large interlayer spacing (12.75 Å), and moderately high $T_c$ (3.1 K). Its symbol size, corresponding to an Ising SOC strength of 3.58, is markedly larger than that of most comparable systems.

$(BaS)_{1/3}TaS_2$ exhibits a $T_c$ of 3.1 K, more than three times that of pristine 2H-$TaS_2$, paired with an exceptionally high $B_{c2}^{/\!/}$ of 20.44 T, one of the highest values reported in $TaS_2$-based superconductors. This combination demonstrates the synergistic interplay between charge-doping-enhanced $T_c$ and Ising-SOC-protected high in-plane critical fields.

Figure 6(b) correlates the spacing between adjacent $(MX_2)_n$ unit with the superconducting anisotropy $B_{c2}^{/\!/}/B_{c2}^{\perp}$. A clear positive trend emerges: larger interlayer distances suppress the interlayer Josephson coupling $t^{\perp}$, thereby enhancing anisotropy—a structural prerequisite for pronounced 2D superconductivity. Parent TMDs such as 2H-$TaS_2$ (spacing $d \approx 6.04$ Å, anisotropy $\approx 6$) and 2H-$TaSe_2$ ($d \approx 6.35$ Å, anisotropy $\approx 3$) display low anisotropy due to strong interlayer coupling. By contrast, intercalated systems with expanded spacings, e.g., $BaCl_{0.75}TaSe_2$ ($d$ = 12.32 Å, anisotropy $\approx 150$)[19] and $(LaSe)_{1.14}NbSe_2$ ($d$ = 12.04 Å, anisotropy $\approx 32$),[25] achieve high anisotropy, but often at the expense of drastically reduced $T_c$ (1.75 K and 1.3 K, respectively). Our $(BaS)_{1/3}TaS_2$ features an interlayer spacing of 12.75 Å, more than double that of the parent compound, and an anisotropy of 34, substantially exceeding those of similarly spaced analogues such as $(PbS)_{1.13}TaS_2$ ($d$ = 12.0 Å, anisotropy = 13.16)[36] and $BaCl_{0.75}TaS_2$ ($d$ = 12.03 Å, anisotropy

= 16.1)[19]. Importantly, it retains a high $T_c$ of 3.1 K, thereby overcoming the conventional trade-off between large spacing/high anisotropy and high $T_c$.

Figure 6(c) provides a 3D visualization of the interplay among anisotropy ($x$-axis), interlayer spacing ($y$-axis), and $T_c$ ($z$-axis), with scatter sizes proportional to the Ising SOC strength ($B_{c2}^{\parallel}/B_p$). This representation synthesizes structural, electronic, and superconducting characteristics. Most systems cluster either in the "low anisotropy–small spacing–low $T_c$" region (e.g., parent TMDs) or the "high anisotropy–large spacing–low $T_c$" region (e.g., many organic/inorganic-intercalated compounds). Recent reports on $(PbSe)_{1.14}(NbSe_2)_3$ and $(SnS)_{1.15}TaS_2$ thin flakes have also demonstrated high $T_c$ with Ising protection.[43-44] Notably, however, these systems are all realized in artificially fabricated thin-flake devices, which inherently hinders the accurate elucidation of their intrinsic bulk material properties. The data point for $(BaS)_{1/3}TaS_2$ stands apart, occupying a region of concurrently high anisotropy (34), large spacing (12.75 Å), and moderate-high $T_c$ (3.1 K). Its corresponding Ising SOC strength of 3.58 markedly exceeds that of typical analogues (e.g., 2.02 for $Na_{0.1}TaS_2$; 1.33 for $Pb_{1/3}TaS_2$)[28,30], underscoring the coexistence of strong 2D confinement, elevated $T_c$, and pronounced Ising protection.

## CONLUSION

Collectively, these experimental results unambiguously validate the efficacy of the Ba-S-S-Ba chain intercalation strategy for engineering the superconducting and structural properties of 2H-$TaS_2$. This unique structural design exerts a dual synergistic effect on the host lattice: on one hand, it suppresses the competing CDW order, which relieves the competition for electronic states at the Fermi level and thus enhances Cooper pair formation to boost the superconducting transition temperature $T_c$; on the other hand, the one-dimensional chain-like geometry of the intercalant drastically expands the interbilayer spacing and efficiently suppresses interlayer electronic coupling, which stabilizes robust bulk 2D Ising superconductivity with an ultrahigh in-plane upper critical field in $(BaS)_{1/3}TaS_2$. In doing so, our work resolves the long-standing trade-off between achieving a high $T_c$ and maintaining strong 2D electronic character in intercalated TMDs, and demonstrates that rational structural design can disentangle these inherently competing material requirements to realize well-defined 2D Ising superconductivity in macroscopic bulk single crystals.

More broadly, this chain-intercalation-driven symmetry engineering approach establishes a generalizable and versatile route to tailor the electronic dimensionality and superconducting properties of layered quantum materials. By precisely manipulating interlayer coupling and symmetry breaking via atomic-level structural design, this strategy opens up new avenues for the deliberate synthesis of high-performance low-dimensional superconducting systems. It not only enriches the fundamental research landscape of Ising superconductivity and topological non-trivial superconducting states but also lays a structural and material foundation for the development of next-generation low-dimensional superconducting devices, thus advancing both the basic science and applied potential of low-dimensional quantum matter.

## ASSOCIATED CONTENT

### Supporting Information

The Supporting Information is available free of charge on the ACS Publications website at DOI. It includes the details of structure refinement, the analysis of the upper critical field, the superconducting gap symmetry, and the demagnetization factor.

## AUTHOR INFORMATION

**Corresponding Author**

*lmchen@zua.edu.cn; *zzyang@phy.ecnu.edu.cn; *guoyf@shanghaitech.edu.cn

**Author Contributions**

Y.F.G. conceived the project. Z.Y.Z and X.Q.L. prepared the samples and performed the crystal characterizations with the help from N.Y. X.Q.L. and C.X. measured the electrical transport properties with the help from Z.Y., Z.Y.L., W.X., J.W.L. and X.W.  H.N.W. carried out the STEM measurements supervised by K.Q. and Z.Z.Y.  X.Q.L., L.M.C. and Y.F.G. wrote the wrote the paper with contributions from all coauthors. †These authors contributed equally.

**Notes**

The authors declare no competing financial interest.

**ACKNOWLEDGMENTS**

The authors acknowledge the National Key R&D Program of China (Grants No. 2023YFA1406100 and 2024YFA1408400). Y.F.G. acknowledges the support by the Key Laboratory of Aeronautical Materials and Application Technology (ZHKF-250103), Science and Technology Commission of Shanghai Municipality (25DZ3008200) and the open research fund of Beijing National Laboratory for Condensed Matter Physics (2023BNLCMPKF002). Z. Z. Yang acknowledges the National Natural Science Foundation of China (Grant No. 52572129) and the Shanghai Committee of Science and Technology (Grant No. 24JD1401200). The authors also thank the support from Analytical Instrumentation Center (#SPST-AIC10112914) and the Double First-Class Initiative Fund of ShanghaiTech University.

## REFERENCES

(1) Tinkham, M. *Introduction to Superconductivity*; 2nd ed.; McGraw-Hill: New York, NY, USA, **1996**.

(2) Clogston, A. M. Upper limit for the critical field in hard superconductors. *Phys. Rev. Lett.* **1962**, 9, 266–267.

(3) Lu, J. M.; Zheliuk, O.; Leermakers, I.; Yuan, N. F. Q.; Zeitler, U.; Law, K. T.; Ye, J. T. Evidence for two-dimensional Ising superconductivity in gated $MoS_2$. *Science* **2015**, 350, 1353–1357.

(4) Xi, X.; Wang, Z.; Zhao, W.; Park, J.-H.; Law, K. T.; Berger, H.; Forró, L.; Shan, J.; Mak, K. F. Ising pairing in superconducting $NbSe_2$ atomic layers. *Nat. Phys.* **2016**, 12, 139–143.

(5) Saito, Y.; Nakamura, Y.; Bahramy, M. S.; Kohama, Y.; Ye, J.; Kasahara, Y.; Nakagawa, Y.; Onga, M.; Tokunaga, M.; Nojima, T.; Yanase, Y.; Iwasa, Y. Superconductivity protected by spin-valley locking in ion-gated $MoS_2$. *Nat. Phys.* **2016**, 12, 144–149.

(6) de la Barrera, S. C.; Sinko, M. R.; Gopalan, M. R.; Sivadas, R.; Seyler, K. L.; Watanabe, K.; Taniguchi, T.; Tsen, A.W.; Xu, X.; Xiao, D.; Hunt, B. M. Tuning Ising superconductivity with layer and spin-orbit coupling in two-dimensional transition-metal dichalcogenides. *Nat. Commun.* **2018**, 9, 1427.

(7) Cui, J.; Li, P.; Zhou, J.; He, W.-Y.; Huang, X.; Yi, J.; Fan, J.; Ji, Z.; Jing, X.; Qu, F.; Cheng, Z. G.; Yang, C.; Lu, L.; Suenaga, K.; Liu, J.; Law, K. T.; Lin, J.; Liu, Z.; Liu, G. Transport evidence of asymmetric spin–orbit coupling in few-layer superconducting $1T_d$-$MoTe_2$. *Nat. Commun.* **2019**, 10, 2044.

(8) Xiao, D.; Liu, G.-B.; Feng, W.; Xu, X.; Yao, W. Coupled spin and valley physics in monolayers of $MoS_2$ and other group-VI dichalcogenides. *Phys. Rev. Lett.* **2011**, 108, 196802.

(9) Kormányos, A.; Zólyomi, V.; Drummond, N. D.; Rakyta, P.; Burkard, G.; Fal'ko, V. I. Monolayer $MoS_2$: Trigonal warping, the Γ valley, and spin-orbit coupling effects. *Phys. Rev. B* **2013**, 88, 045416.

(10) Sohn, E.; Xi, X.; He, W.-Y.; Jiang, S.; Wang, Z.; Kang, K.; Park, J.-H.; Berger, H.; Forró, L.; Law, K. T.; Shan, J.; Mak, K. F. An unusual continuous paramagnetic-limited superconducting phase transition in 2D $NbSe_2$. *Nat. Mater.* **2018**, 17, 504.

(11) Tsen, A. W.; Hunt, B.; Kim, Y. D.; Yuan, Z. J.; Jia, S.; Cava, R. J.; Hone, J.; Kim, P.; Dean, C. R.; Pasupathy, A. N. Nature of the quantum metal in a two-dimensional crystalline superconductor. *Nat. Phys.* **2016**, 12, 208.

(12) Zhou, B. T.; Yuan, N. F. Q.; Jiang, H.-L.; Law, K. T. Ising superconductivity and Majorana fermions in transition-metal dichalcogenides. *Phys. Rev. B* **2016**, 93, 180501.

(13) Devarakonda, A.; Inoue, H.; Fang, S.; Ozsoy-Keskinbora, C.; Suzuki, T.; Kriener, E.; Fu, L.; Kaxiras, E.; Bell, D. C.; Checkelsky, J. G. Clean 2D superconductivity in a bulk van der Waals superlattice. *Science* **2020**, 370, 231.

(14) Wan, P.; Zheliuk, O.; Yuan, N. F. Q.; Peng, X.; Zhang, L.; Liang, M.; Zeitler, U.; Wiedmann, S.; Hussey, N. E.; Palstra, T. T. M.; Ye, J. Orbital Fulde–Ferrell–Larkin–Ovchinnikov state in an Ising superconductor. *Nature* **2023**, 619, 46–51.

(15) Zhang, H.; Rousuli, A.; Zhang, K.; Luo, L.; Guo, C.; Cong, X.; Lin, Z.; Bao, C.; Zhang, H.; Xu, S.; Feng, R.; Shen, S.; Zhao, K.; Yao, W.; Wu, Y.; Ji, S.; Chen, X.; Tan, P.; Xue, Q.-K.; Xu,

Y.; Duan, W.; Yu, P.; Zhou, S. Tailored Ising superconductivity in intercalated bulk $NbSe_2$. *Nat. Phys.* **2022**, 18, 1425.

(16) Sun, R.; Deng, J.; Wu, X.; Hao, M.; Ma, K.; Ma, Y.; Zhao, C.; Meng, D.; Ji, X.; Ding, Y.; Pang, Y.; Qian, X.; Yang, R.; Li, G.; Li, Z.; Dai, L.; Ying, T.; zhao, H.; Du, S.; Li, G.; Jin, S.; Chen, X. High anisotropy in electrical and thermal conductivity through the design of aerogel-like superlattice $(NaOH)_{0.5}NbSe_2$. *Nat. Commun.* **2023**, 14, 6689.

(17) Samuely, P.; Szabó, P.; Kačmarčík, J.; Meerschaut, A.; Cario, L.; Jansen, A. G. M.; Cren, T.; Kuzmiak, M.; Šofranko, O.; Samuely, T. Extreme in-plane upper critical magnetic fields of heavily doped quasi-two-dimensional transition metal dichalcogenides. *Phys. Rev. B* **2021**, 104, 224507.

(18) Yang, X.; Yu, T.; Xu, C.; Wang, J.; Hu, W.; Xu, Z.; Wang, T.; Zhang, C.; Ren, Z.; Xu, Z.-a.; Hirayama, M.; Arita, R.; Lin, X. Anisotropic superconductivity in topological crystalline metal $Pb_{1/3}TaS_2$ with multiple Dirac fermions. *Phys. Rev. B* **2021**, 104, 035157.

(19) Shi, M.; Fan, K.; Li, H.; Pan, S.; Cai, J.; Zhang, N.; Li, H.; Wu, T.; Zhang, J.; Xi, C.; Xiang, Z.; Chen, X. Two-dimensional superconductivity and anomalous vortex dissipation in newly-discovered transition metal dichalcogenide-based superlattices. *J. Am. Chem. Soc*. **2024**, 146, 33413–33422.

(20) Devarakonda, A.; Chen, A.; Fang, S.; Graf, D.; Kriener, M.; Akey, A. J.; Bell, D. C.; Suzuki, T.; Checkelsky, J. G. Evidence of striped electronic phases in a structurally modulated superlattice. *Nature* **2024**, 631, 526.

(21) Ma, K.; Jin, S.; Meng, F.; Zhang, Q.; Sun, R.; Deng, J.; Chen, L.; Gu, L.; Li, G.; Zhang, Z. Two-dimensional superconductivity in a bulk superlattice van der Waals material $Ba_6Nb_{11}Se_{28}$. *Phys. Rev. Mater.* **2022**, 6 (4), 044806.

(22) Zheng, B.; Zhang, X.; Wang, K.; Li, R.; Cao, J.; Wang, C.; Tan, H.; Li, Z.; Lin, B.; Li, P.; Xi, C.; Zhang, J.; Lu, Y.; Zhu, W.; Liu, Z.; Yang, S. A.; Li, L.-J.; Liu, F.; Xiang, B. 3D Ising Superconductivity in As-Grown Sn Intercalated $TaSe_2$ Crystal. *Nano Lett.* **2025**, 25 (12), 4895.

(23) Li, L. J.; Zhu, X. D.; Sun, Y. P.; Lei, H. C.; Wang, B. S.; Zhang, S. B.; Zhu, X. B.; Yang, Z. R.; Song, W. H. Superconductivity of Ni-doping 2H-$TaS_2$. *Physica C: Superconductivity* **2010**, 470, 313.

(24) Agarwal, T.; Patra, C.; Kataria, A.; Chowdhury, R. R.; Singh, R. P. Quasi-two-dimensional anisotropic superconductivity in Li-intercalated 2H-$TaS_2$. *Phys. Rev. B* **2023**, 107 (17), 174509.

(25) Zullo, L.; Setnikar, G.; Pawbake, A.; Cren, T.; Brun, C.; Cordiez, J.; Sasaki, S.; Cario, L.; Marini, G.; Calandra, M.; Méasson, M.-A. Charge density wave collapse of $NbSe_2$ in the $(LaSe)_{1.14}(NbSe_2)_2$ misfit layer compound. *Phys. Rev. B* **2024**, 110, 075430.

(26) Hsu, Y.-T.; Vaezi, A.; Fischer, M. H.; Kim, E.-A. Topological superconductivity in monolayer transition metal dichalcogenides. *Nat. Commun.* **2017**, 8, 14985.

(27) Li, Y. W.; Zheng, H. J.; Fang, Y. Q.; Zhang, D. Q.; Chen, Y. J.; Chen, C.; Liang, A. J.; Shi, W. J.; Pei, D.; Xu, L. X.; Liu, S.; Pan, J.; Lu, D. H.; Hashimoto, M.; Barinov, A.; Jung, S. W.; Cacho, C.; Wang, M. X.; He, Y.; Fu, L.; Zhang, H. J.; Huang, F. Q.; Yang, L. X.; Liu, Z. K.; Chen, Y. L. Observation of topological superconductivity in a stoichiometric transition metal dichalcogenide 2M-$WS_2$. *Nat. Commun.* **2021**, 12, 2874.

(28) Fang, L.; Wang, Y.; Zou, P. Y.; Tang, L.; Xu, Z.; Chen, H.; Dong, C.; Shan, L.; Wen, H. H. Fabrication and superconductivity of $Na_xTaS_2$ crystals. *Phys. Rev. B* **2005**, 72, 014534.

(29) Zhu, X.; Sun, Y.; Zhang, S.; Wang, J.; Zou, L.; DeLong, L. E.; Zhu, X.; Luo, X.; Wang, B.; Li, G.; Yang, Z.; Song, W. Anisotropic intermediate coupling superconductivity in $Cu_{0.03}TaS_2$. *J. Phys.: Condens. Matter* **2009**, 21, 145701.

(30) Li, Y.; Wu, Z.; Zhou, J.; Bu, K.; Xu, C.; Qiao, L.; Li, M.; Bai, H.; Ma, J.; Tao, Q.; Cao, C.; Yin, Y.; Xu, Z.-A. Enhanced anisotropic superconductivity in the topological nodal-line semimetal $In_xTaS_2$. *Phys. Rev. B* **2020**, 102, 224503.

(31) Silber, I.; Mathimalar, S.; Mangel, I.; Nayak, A. K.; Green, O.; Avraham, N.; Beidenkopf, H.; Feldman, I.; Kanigel, A.; Klein, A.; Goldstein, M.; Banerjee, A.; Sela, E.; Dagan, Y. Two-component nematic superconductivity in $4H_b$-$TaS_2$. *Nat. Commun.* **2024**, 15, 824.

(32) Zheng, B.; Feng, X.; Liu, B.; Liu, Z.; Wang, S.; Zhang, Y.; Ma, X.; Luo, Y.; Wang, C.; Li, R.; Zhang, Z.; Cui, S.; Lu, Y.; Sun, Z.; He, J.; Yang, S. A.; Xiang, B. The Coexistence of Superconductivity and Topological Order in Van der Waals $InNbS_2$. *Small* **2024**, 20, 2305909.

(33) Qiu, D.; Zou, Y.; Yang, C.; Zheng, D.; Zhang, C.; Zhang, D.; Wu, Y.; Rao, G.; Li, P.; Zhou, Y.; Jian, X.; Wei, H.; Cheng, Z.; Zhang, X.; Zhang, Y.; Liu, H.; Qi, J.; Li, Y.; Xiong, J. Quantum fluctuations-driven melting transitions in two-dimensional superconductors. *Phys. Rev. Research* **2025**, 7, 033025.

(34) Huang, X.; Jia, L.; Song, X.; Chen, Y.; Song, Y.; Yang, K.; Guo, J.-g.; Huang, Y.; Liu, L.; Wang, Y. Observation of two-dimensional type-II superconductivity in bulk 3R-$TaSe_2$ by scanning tunneling spectroscopy. *J. Phys. Chem. Lett.* **2023**, 14, 7235–7240.

(35) Agarwal, T.; Patra, C.; Manna, P.; Srivastava, S.; Mishra, P.; Sharma, S.; Singh, R. P. Anomalous magnetotransport in the superconducting architecturally misfit layered system $(PbS)_{1.13}TaS_2$. *Phys. Rev. B* **2025**, 112, 014501.

(36) Smith, T. F.; Shelton, R. N.; Schwall, R. E. Superconductivity of $TaS_{2-x}Se_x$ at high pressure. *J. Phys. F: Metal Phys.* **1975**, 5, 1713.

(37) Varma, C. M.; Littlewood, P. B.; Schmitt-Rink, S.; Abrahams, E.; Ruckenstein, A. E. Phenomenology of the normal state of Cu-O high-temperature superconductors. *Phys. Rev. Lett.* **1989**, 63, 1996.

(38) Onabe, K.; Naito, M.; Tanaka, S. Anisotropy of Upper Critical Field in Superconducting 2H-$NbS_2$. *J. Phys. Soc. Jpn.* **1978**, 45, 50–58.

(39) Zhou, M.; Ni, S.; Shi, Y.; Chen, L.; Yi, J.; Gu, Y.; Liu, Q.; Ruan, B.; Yang, Q.; Yu, J.; Yang, Q. Nodeless two-gap superconductivity revealed by low-temperature specific heat and lower critical field in 4Ha-$NbSe_2$. *Phys. Rev. B* **2025**, 111, 024513.

(40) Yu, L.; Mi, M.; Xiao, H.; Wang, S.; Sun, Y.; Lyu, B.; Bai, L.; Shen, B.; Liu, M.; Wang, S.; Wang, Y. Intercalation-Induced Monolayer Behavior in Bulk $NbSe_2$. *ACS Appl. Mater. Interfaces* **2024**, 16 (43), 59049-59055

(41) Wu, J.; Wu, J.; Wang, C.; Hu, G.; Feng, Y.; Ma, X.; Li, R.; Si, C.; Lu, Y.; Li, L.; Xiang, B.; Zhang, Z. Enhanced Ising Superconductivity and Emergent Ferromagnetism Coexisting in Bimolecule-Intercalated Bulk $TaS_2$. *Adv. Mater.* **2025** 37, 2503794.

(42) Margineda, D.; Álvarez-García, C.; Tezze, D.; Gerivani, S.; Caldevilla-Asenjo, D.; Furqan, M.; Rivilla, I.; Casanova, F.; Arenal, R.; Artacho, A.; Hueso, L.; Gobbi, M. Degenerate Monolayer

Ising Superconductors via Chiral-Achiral Molecule Intercalation. *Adv. Funct. Mater.* **2025**, 36(25), e27724.

(43) Itahashi, Y.M.; Nohara, Y.; Chazono, M.; Matsuoka, H.; Arioka, K.; Nomoto, T.; Kohama, Y.; Yanase, Y.; Iwasa, Y.; Kobayashi, K. Misfit layered superconductor $(PbSe)_{1.14}(NbSe_2)_3$ with possible layer-selective FFLO state. *Nat. Commun.* **2025**, 16, 7022.

(44) Li, Z.; Lyu, P.; Chen, Z.; Guan, D.; Yu, S.; Zhao, J.; Huang, P.; Zhou, X.; Qiu, Z.; Fang, H.; Hashimoto, M.; Lu, D.; Song, F.; Loh, K. P.; Zheng, Y.; Shen, Z.-X.; Novoselov, K. S.; Lu, J. Beyond Conventional Charge Density Wave for Strongly Enhanced 2D Superconductivity in 1H-$TaS_2$ Superlattices. *Adv. Mater.* 2**024**, 36, 2312341.

BRIEFS: The unique chain-like intercalation in $(BaS)_{1/3}TaS_2$ overcomes the conventional trade-off between high Tc and robust two-dimensionality, achieving bulk Ising superconductivity characterized by a high in-plane upper critical field and strong anisotropy. This system establishes a novel intercalation strategy for designing high-performance low-dimensional superconductors.

**TOC:**

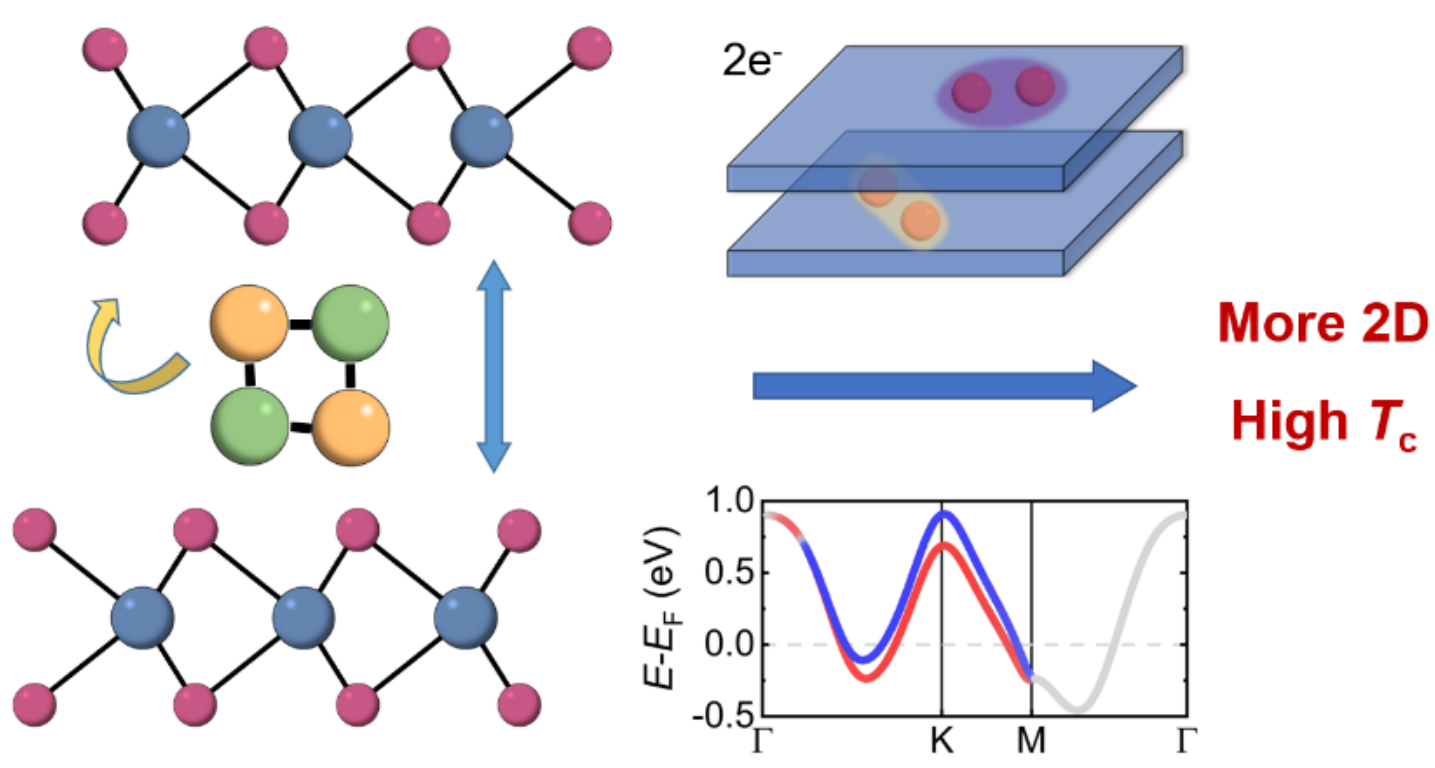

# SI

## 1. Upper critical field and two-dimensional superconductivity

The temperature dependence of upper critical field $B_{c2}$ of $(BaS)_{1/3}TaS_2$ is modelled using the Ginzburg-Landau (G-L) framework for both three-dimensional (3D) and 2D superconductivity.

For 3D superconductivity:

$$B_{c2} = \frac{\Phi_0}{2\pi\xi_0^2}\left(1-\frac{T}{T_c}\right). \tag{1}$$

For 2D superconductivity:

$$B_{c2} = \frac{\sqrt{3}\Phi_0}{\pi\xi_0 d_{sc}}(1-T/T_c)^{\frac{1}{2}}, \tag{2}$$

where $\Phi_0$ represents the flux quantum, $\xi_0$ is the zero-temperature GL coherence length, $d_{sc}$ denotes the 2D superconducting thickness.

In $(BaS)_{1/3}TaS_2$, vacancies in the Ba-S layer may introduce strong spin scattering, significantly affecting Cooper pair depairing under magnetic fields. The Werthamer–Helfand–Hohenberg (WHH) theory provides a framework for understanding pair-breaking mechanisms in type-II superconductors, incorporating contributions from both Pauli paramagnetism and spin–orbit scattering (SOS). The WHH model assumes a dirty-limit scenario where the SOS rate is much weaker than the momentum-scattering rate and orbital diamagnetism is negligible. Under these conditions, pair-breaking is quantified by two relaxation times: the spin-independent scattering time ($\tau_1$) and the spin–orbit scattering time ($\tau_2$). The total relaxation rate ($\tau^{-1} = \tau_1^{-1} + \tau_2^{-1}$) is dominated by momentum scattering ($\tau << \tau_1$).

The WHH theory predicts a pronounced upward curvature in $B_{c2}(T)$, reflecting a strong enhancement of the zero-temperature critical field $B_{c2}(0)$ due to SOS. SOS randomizes spin orientation, counteracting the alignment of spins in Cooper pairs by an external magnetic field. This suppression of Pauli paramagnetism allows $B_{c2}$ to exceed the conventional Pauli limit.[1,2] The temperature dependence of $B_{c2}^{/\!/}(T)$ is described by:

$$\ln\left(\frac{1}{t}\right) = \left(\frac{1}{2}+\frac{i\lambda_{\mathrm{so}}}{4\gamma}\right)\psi\left(\frac{1}{2}+\frac{\bar{h}+\frac{\lambda_{\mathrm{so}}}{2}+i\gamma}{2t}\right)+\left(\frac{1}{2}-\frac{i\lambda_{\mathrm{so}}}{4\gamma}\right)\psi\left(\frac{1}{2}+\frac{\bar{h}+\frac{\lambda_{\mathrm{so}}}{2}-i\gamma}{2t}\right), \tag{3}$$

where $t = T/T_c$, $\bar{h} = \frac{2e\mu_0 B v_F^2 \tau}{6\pi c k_B T_c}$, and $\gamma = \sqrt{(\alpha\hat{h})^2 - (\lambda_{\mathrm{SO}}/2)^2}$, with the fitting parameters Maki parameter $\alpha = \frac{2\hbar}{2mv_F^2\tau}$ and SOC strength $\lambda_{\mathrm{SO}} = \frac{\hbar}{2\pi k_B T_c \tau_1}$.

The dimensional character of polytype is further elucidated by the angular dependence of the upper critical field, $B_{c2}(\theta)$, shown in Fig. 3. Initially, we model this behavior using the anisotropic G-L model:

$$\left(\frac{B_{c2}(\theta)\sin\theta}{B_{c2}^{\parallel}}\right)^2 + \left(\frac{B_{c2}(\theta)\cos\theta}{B_{c2}^{\perp}}\right)^2 = 1, \tag{4}$$

where $B_{c2}^{/\!/}$ and $B_{c2}^{\perp}$ represent the critical fields for magnetic fields oriented parallel ($\theta = 90^{\circ}$) and perpendicular ($\theta = 0^{\circ}$) to the layers, respectively. However, given the system's pronounced 2D nature, we also apply the 2D Tinkham model:

$$\left(\frac{B_{c2}(\theta)\sin\theta}{B_{c2}^{\parallel}}\right)^2 + \left|\frac{B_{c2}(\theta)\cos\theta}{B_{c2}^{\perp}}\right| = 1, \tag{5}$$

The two models diverge significantly as the field approaches the layer plane ($\theta \sim 90^{\circ}$), with the Tinkham model predicting a sharper enhancement of $B_{c2}$within the plane—a feature corroborated by our experimental data.

## 2. Superconducting Gap Symmetry Analysis

To elucidate the superconducting gap structure, we employ the BCS framework to model the electronic specific heat:

$$\gamma_e - \gamma_n = \frac{4N(E_F)}{k_B T^3} \int_0^{+\infty} \int_0^{2\pi} \frac{e^{\zeta/k_B T}}{\left(1+e^{\zeta/k_B T}\right)^2} \times \left[\varepsilon^2 + \Delta^2(T,\theta) - \frac{T}{2}\frac{d\Delta^2(\theta,T)}{dT}\right] d\theta\, d\varepsilon, \quad (6)$$

where $\zeta = \sqrt{\varepsilon^2 + \Delta^2(T,\theta)}$. We use the isotropic *s*-wave ($\Delta(T, \theta) = \Delta_0(T)$) to estimate the superconducting gap. The gap amplitude follows a temperature dependence that saturates below 3.1 K, well described by:

$$\Delta(T) = \Delta_0 \tanh\left[1.74\sqrt{\frac{T_c}{T} - 1}\right], \quad (7)$$

## 3. Demagnetization Factor

To calculated the superconducting volume from magnetization, we consider the demagnetization fact in Meissner state. For our cuboid sample with dimensions $a^*$ ~ 0.6 mm, $b^*$ ~ 0.4 mm, and $c^*$ ~ 0.05 mm, the demagnetization factor $N$ is determined by:

$$N^{-1} = 1 + \frac{3}{4}\frac{c^*}{a^*}\left(1+\frac{a^*}{b^*}\right), \quad (8)$$

Yielding $N_{B/\!/c} = 0.865$. The internal magnetic field $H_i$ follows:

$$H_i = \frac{H_0}{1+\chi N}, \quad (9)$$

## 4. Crystal Data and Structure Refinement

**Table S1.** Crystallographic data of $(BaS)_{1/3}TaS_2$ at 298 K by the structural refinement of the single-crystal XRD.

| | |
|---|---|
| Chemical formula | $(BaS)_{1/3}TaS_2$ |
| Formula weight | 295.884 g/mol |
| Temperature | 298.00 K |
| Crystal system | trigonal |
| Space group | $P$-3$m$1 |
| | $a$ = 3.31790(10) Å, $\alpha$ = 90° |
| Unit-cell dimensions | $b$ = 3.31790(10) Å, $\beta$= 90° |
| | $c$ = 18.7414(15) Å, $\gamma$= 120° |
| Volume | 178.67 (3) $Å^3$ |
| Z | 2 |
| Density | 5.610 $g/cm^3$ |
| Absorption coefficient μ | 35.456 $mm^{-1}$ |
| $F$ (000) | 258 |
| Crystal size | 0.5 × 0.5 × 0.05 $mm^3$ |
| Radiation | Mo $K\alpha$ ($\lambda$ = 0.71073 Å) |
| 2Θ range for data collection | 5.34° to 52.62° |
| Index ranges | $-4 \leq h \leq 4$, $-3 \leq k \leq 4$, $-23 \leq l \leq 23$ |
| Reflections collected | 1937 |
| Independent reflections | 190 |
| Data/restraints/parameters | 190/0/15 |
| Goodness-of-fit on $F^2$ | 1.156 |
| Final $R$ indexes [$I >= 2\sigma (I)$] | $R_1$ = 0.0653, $wR_2$ = 0.1860 |
| Final $R$ indexes [all data] | $R_1$ = 0.0653, $wR_2$ = 0.1859 |
| Largest diff. peak/hole | 3.69 e $Å^{-3}$/-2.824 e $Å^{-3}$ |

| Label | x | y | z | Occupancy | $U_{eq}$* |
|---|---|---|---|---|---|
| Ta 1 | 1.00000 | 0.00000 | 0.8401(1) | 1.00000 | 0.027(1) |
| S 2 | 0.66667 | 0.33333 | 0.9216(5) | 1.00000 | 0.029(2) |
| S 3 | 0.66667 | 0.33333 | 0.7563(5) | 1.00000 | 0.029(2) |
| Ba 4 | 1.00000 | 0.00000 | 0.6115(6) | 0.33333 | 0.087(5) |
| S 5 | 1.00000 | 0.00000 | 0.4751(2) | 0.33333 | 0.028(7) |

*$U_{eq}$ is defined as 1/3 of the trace of the orthogonalized $U_{ij}$ tensor. The Anisotropic displacement factor exponent takes the form: $-2\pi^2[h2a * 2U_{11} + ... + 2hka * b * U_{12}]$

| Label | $U_{11}$ | $U_{22}$ | $U_{33}$ | $U_{12}$ | $U_{13}$ | $U_{23}$ |
|---|---|---|---|---|---|---|
| Ta 1 | 0.025(1) | 0.025(1) | 0.032(1) | 0 | 0 | 0.0125(5) |
| S 2 | 0.026(3) | 0.026(3) | 0.033(4) | 0 | 0 | 0.0132(2) |
| S 3 | 0.027(3) | 0.027(3) | 0.033(4) | 0 | 0 | 0.0135(2) |
| Ba 4 | 0.109(8) | 0.109(8) | 0.042(5) | 0 | 0 | 0.0550(4) |
| S 5 | 0.129(2) | 0.129(2) | 0.037(7) | 0 | 0 | 0.0650(9) |

## 5. Comparison of resistivity anisotropy between $(BaS)_{1/3}TaS_2$ and 2H-$TaS_2$

To further highlight the exceptional quasi-two-dimensional electronic character of our $(BaS)_{1/3}TaS_2$ system, we added a detailed comparison of the resistivity anisotropy between $(BaS)_{1/3}TaS_2$ and pristine 2H-$TaS_2$ (Figures S1(a) and 1(b), respectively). As shown in Figure S1, pristine 2H-$TaS_2$ exhibits a moderate resistivity anisotropy ratio $\rho_{zz}/\rho_{xx}$ in the range of 10–50 (Figure S1(c)). In contrast, our $(BaS)_{1/3}TaS_2$ system achieves an ultrahigh anisotropy ratio exceeding $4\times10^3$ at low temperatures (Figure S1(d)). This striking difference confirms that the unique Ba-S-S-Ba chain intercalation induces an extreme quasi-two-dimensional electronic character in the bulk material—without compromising its structural or physical integrity.

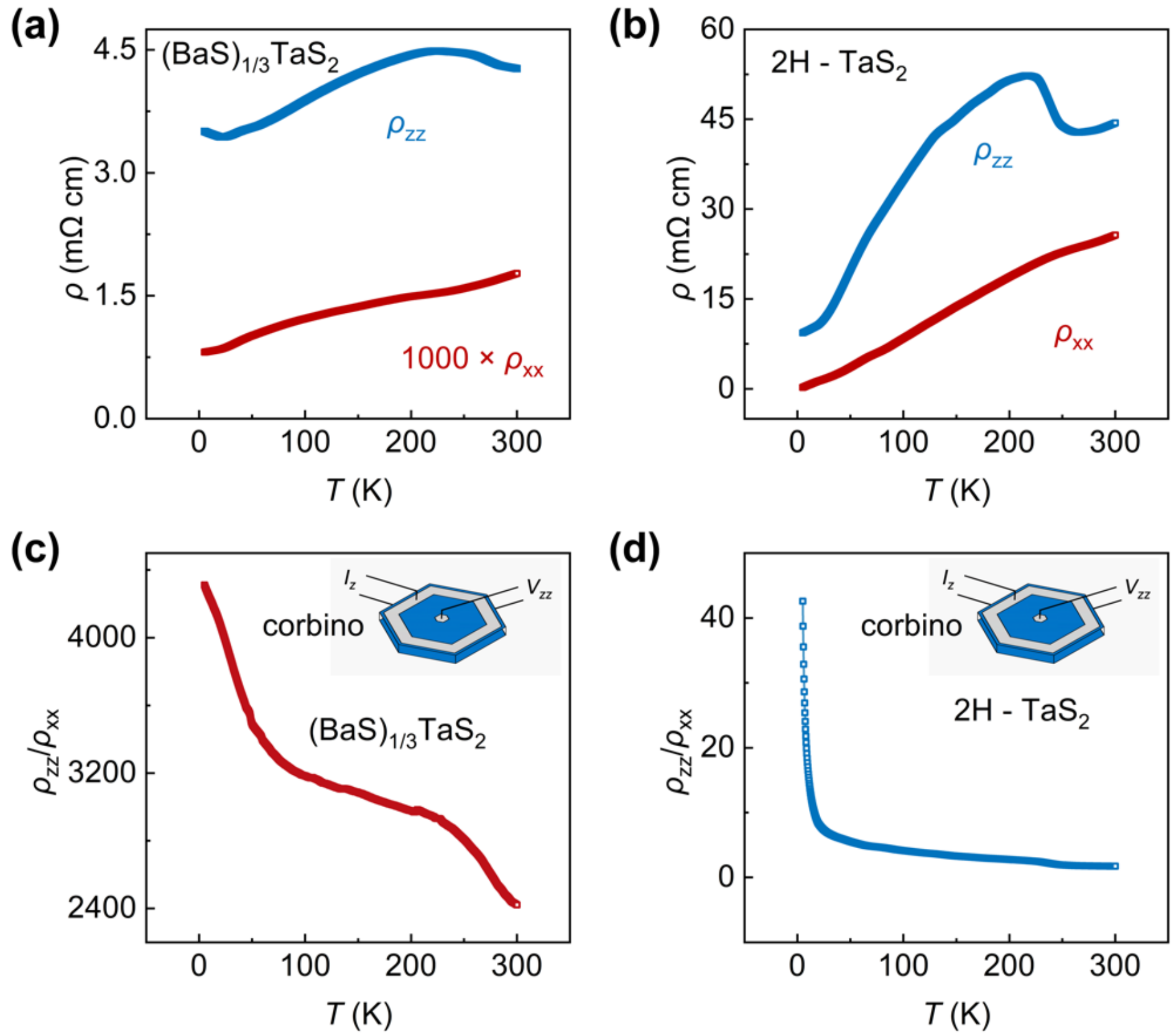

**Figure S1**. Comparison of the resistivity anisotropy $\rho_{zz}/\rho_{xx}$ between $(BaS)_{1/3}TaS_2$ and 2H-$TaS_2$ using the Corbino geometry.

## REFERENCES

(1) Wan, P., Zheliuk, O., Yuan, N.F.Q. et al. Orbital Fulde–Ferrell–Larkin–Ovchinnikov state in an Ising superconductor. *Nature* **2023**, 619, 46–51.

(2) Zhang, H. X.; Rousuli, A.; Zhang, K.; Luo, L.; Guo, C.; Cong, X.; Lin, Z.; Bao, C.; Zhang, H.; Xu, S.; Feng, R.; Shen, S.; Zhao, K.; Yao, W.; Wu, Y.; Ji, S.; Chen, X.; Tan, P.; Xue, Q.-K.; Xu, Y.; Duan, W.; Yu, P.; Zhou, S. Tailored Ising superconductivity in intercalated bulk $NbSe_2$. *Nat. Phys.* **2022**, 18, 1425.